\documentclass[12pt]{article}

\usepackage{amsmath,amssymb,amscd,amsbsy,amsgen,amsopn,amstext,
amsxtra}

\usepackage[dvips]{graphicx}

\begin{document}

\begin{flushright}
NUP-A-2005-2
\end{flushright}

\vskip 0.5 truecm

\begin{center}
{\Large{\bf Second quantized formulation of geometric phases}}
\end{center}
\vskip .5 truecm
\centerline{\bf Shinichi Deguchi {\rm and} Kazuo Fujikawa }
\vskip .4 truecm
\centerline {\it Institute of Quantum Science, College of 
Science and Technology}
\centerline {\it Nihon University, Chiyoda-ku, Tokyo 101-8308, 
Japan}
\vskip 0.5 truecm

\makeatletter
\@addtoreset{equation}{section}
\def\theequation{\thesection.\arabic{equation}}
\makeatother

\begin{abstract}
The level crossing problem and associated geometric terms are
neatly formulated by the second quantized formulation. This 
formulation exhibits a hidden local gauge symmetry related to 
the arbitrariness of the phase choice of the complete 
orthonormal basis set. By using 
this second quantized formulation, which does not assume 
adiabatic approximation, a convenient exact formula for the 
geometric 
terms including off-diagonal geometric terms is derived. The 
analysis of geometric phases is then reduced to a simple 
diagonalization of the Hamiltonian, and it is 
analyzed both in the operator and path integral formulations. 
If one diagonalizes the geometric terms in the infinitesimal 
neighborhood of level crossing, the geometric phases become 
trivial (and thus no monopole singularity) for arbitrarily large
 but finite time interval $T$. The integrability of 
Schr\"{o}dinger equation and the appearance of the seemingly
 non-integrable phases are thus consistent. The topological 
proof of the Longuet-Higgins' phase-change rule, for example, 
fails in the practical Born-Oppenheimer approximation where a 
large but finite ratio of
 two time scales is involved and $T$ is identified with the 
period of the slower system. The difference and similarity 
between the geometric phases associated with level crossing and 
the exact topological object such as the Aharonov-Bohm phase 
become clear in the present formulation. A crucial difference 
between the quantum anomaly and the geometric phases is also 
noted.    
\end{abstract}


\section{Introduction}
The geometric phases are usually analyzed in the framework of 
first quantization by using the adiabatic 
approximation~\cite{berry}-\cite{hasegawa2}, though
a non-adiabatic treatment has been considered in, for example, 
\cite{aharonov} and the (non-adiabatic) correction to the 
geometric phases has been analyzed in~\cite{berry2}.
The Hamiltonian, which contains a set of slowly varying external
parameters, has no obvious singularity by itself. But  
a singularity reminiscent of the magnetic monopole is induced 
at the level crossing point, which is controlled by the movement 
of the external parameters, and the associated geometric 
phases appear in the adiabatic approximation. A remarkable fact 
is that the geometric phase factors thus introduced are rather 
universal independently of detailed physical processes. The 
topological properties are considered to be responsible for 
this universal behavior. Also, interesting mathematical ideas 
such as  parallel transport and holonomy are often 
used~\cite{simon} in the framework of adiabatic approximation.

The geometric phases revealed the importance of 
hitherto un-recognized phase factors in the adiabatic 
approximation. It may then be interesting to investigate how 
those phases appear in the exact formulation.
The purpose of the present paper is to formulate the level 
crossing problem by using the second quantization technique, which works both in the path integral and operator formulations.
We thus derive a convenient exact formula for geometric terms, 
including the off-diagonal terms as well as the conventional 
diagonal terms. In this formulation, the analysis of geometric 
phases is reduced to the familiar diagonalization of the  
Hamiltonian. Namely, all the information concerning the 
extra phase factors is contained in the effective Hamiltonian.
In Ref.~\cite{berry2}, this fact that the geometric phases are
interpreted as parts of the Hamiltonian has been noted though
only the diagonal geometric terms have been analyzed in the 
adiabatic picture. Our formulation is more general without 
assuming the adiabatic picture. 

When one diagonalizes the 
Hamiltonian in a very specific limit, one recovers the 
conventional geometric phases defined in the adiabatic 
approximation. One can thus analyze the 
geometric phases in the present formulation without using the 
mathematical notions such as parallel transport and holonomy. 
Instead, a hidden local gauge symmetry plays an important role
in our formulation. If one diagonalizes the 
Hamiltonian in the other extreme limit, namely, in the 
infinitesimal neighborhood of level crossing for any fixed 
finite time interval $T$, one can show that the geometric
phases become trivial and thus no monopole-like
singularity. At the level crossing point, the conventional 
energy eigenvalues become degenerate but the degeneracy is 
lifted if one diagonalizes the geometric terms. 
Since the time interval involved in the practical 
physical processes is always finite, our analysis implies an important change in our understanding of the qualitative aspects of geometric phases. For example, our analysis 
implies that the topological interpretation~\cite{stone, berry} 
of geometric phases such as the topological proof of the 
Longuet-Higgins' phase-change rule~\cite{higgins} fails in the 
practical Born-Oppenheimer approximation where a large but finite
ratio of two time scales is involved and $T$ is identified with
the period of the slower system.

In our analysis, it is important to distinguish the precise
adiabatic approximation, where the time interval $T$ measured in 
units of the shorter time scale is taken to be 
$T\rightarrow\infty$~\cite{simon}, from the practical 
Born-Oppenheimer approximation where a large but finite ratio
of two time scales is involved and the variables with the slower
 time scale are approximately treated as external c-number 
parameters. Our analysis shows that the  
integrability of the Schr\"{o}dinger equation for a regular 
Hamiltonian and the appearance of the seemingly 
``non-integrable phases'' are consistent: To be precise, the 
integrability of the Schr\"{o}dinger equation becomes relevant
when the slowly varying external parameters are promoted to 
the dynamical variables of a more fundamental regular 
Hamiltonian. 

We also clarify the difference between the geometric phases 
associated with level crossing and the exact topological object 
such as the Aharonov-Bohm phase. A crucial difference 
between the quantum anomaly and the geometric phases associated 
with level crossing is also noted.    

The basic idea involved in the present formulation has been  
reported elsewhere~\cite{fujikawa}, and we here present further 
details of the analyses.

\section{Second quantized formulation and geometric phases}

We start with the generic (hermitian) Hamiltonian 
\begin{equation}
\hat{H}=\hat{H}(\hat{\vec{p}},\hat{\vec{x}},X(t))
\end{equation}
for a single particle theory in a slowly varying background 
variable $X(t)=(X_{1}(t),X_{2}(t),...)$.
The path integral for this theory for the time interval
$0\leq t\leq T$ in the second quantized 
formulation is given by 
\begin{eqnarray}
Z&=&\int{\cal D}\psi^{\star}{\cal D}\psi
\exp\{\frac{i}{\hbar}\int_{0}^{T}dtd^{3}x[
\psi^{\star}(t,\vec{x})i\hbar\frac{\partial}{\partial t}
\psi(t,\vec{x})\nonumber\\
&&-\psi^{\star}(t,\vec{x})
\hat{H}(\frac{\hbar}{i}\frac{\partial}{\partial\vec{x}},
\vec{x},X(t))\psi(t,\vec{x})] \}.
\end{eqnarray}
We then define a complete set of eigenfunctions
\begin{eqnarray}
&&\hat{H}(\frac{\hbar}{i}\frac{\partial}{\partial\vec{x}},
\vec{x},X(0))u_{n}(\vec{x},X(0))
=\lambda_{n}u_{n}(\vec{x},X(0)), \nonumber\\
&&\int d^{3}xu_{n}^{\star}(\vec{x},X(0))u_{m}(\vec{x},X(0))=
\delta_{nm},
\end{eqnarray}
and expand 
\begin{eqnarray}
\psi(t,\vec{x})=\sum_{n}a_{n}(t)u_{n}(\vec{x},X(0)).
\end{eqnarray}
We then have
\begin{eqnarray}
{\cal D}\psi^{\star}{\cal D}\psi=\prod_{n}{\cal D}a_{n}^{\star}
{\cal D}a_{n}
\end{eqnarray}
and the path integral is written as 
\begin{eqnarray}
Z&=&\int \prod_{n}{\cal D}a_{n}^{\star}
{\cal D}a_{n}
\exp\{\frac{i}{\hbar}\int_{0}^{T}dt[
\sum_{n}a_{n}^{\star}(t)i\hbar\frac{\partial}{\partial t}
a_{n}(t)\nonumber\\
&&-\sum_{n,m}a_{n}^{\star}(t)E_{nm}(X(t))a_{m}(t)] \}
\end{eqnarray}
where 
\begin{eqnarray}
E_{nm}(X(t))=\int d^{3}x u_{n}^{\star}(\vec{x},X(0))
\hat{H}(\frac{\hbar}{i}\frac{\partial}{\partial\vec{x}},
\vec{x},X(t))u_{m}(\vec{x},X(0)).
\end{eqnarray}

We next perform a unitary transformation
\begin{eqnarray}
a_{n}=\sum_{m}U(X(t))_{nm}b_{m}
\end{eqnarray}
where 
\begin{eqnarray}
U(X(t))_{nm}=\int d^{3}x u^{\star}_{n}(\vec{x},X(0))
v_{m}(\vec{x},X(t))
\end{eqnarray}
with the instantaneous eigenfunctions of the Hamiltonian
\begin{eqnarray}
&&\hat{H}(\frac{\hbar}{i}\frac{\partial}{\partial\vec{x}},
\vec{x},X(t))v_{n}(\vec{x},X(t))
={\cal E}_{n}(X(t))v_{n}(\vec{x},X(t)), \nonumber\\
&&\int d^{3}x v^{\star}_{n}(\vec{x},X(t))v_{m}(\vec{x},X(t))
=\delta_{n,m}.
\end{eqnarray}
We emphasize that $U(X(t))$ is a unit matrix both at $t=0$ and 
$t=T$ if $X(T)=X(0)$, and thus 
\begin{eqnarray}
\{a_{n}\}=\{b_{n}\}
\end{eqnarray}
both at $t=0$ and $t=T$. We take the time $T$ as a period of the 
slowly varying variable $X(t)$.
We can thus re-write the path integral as 
\begin{eqnarray}
&&Z=\int \prod_{n}{\cal D}b_{n}^{\star}{\cal D}b_{n}
\exp\{\frac{i}{\hbar}\int_{0}^{T}dt[
\sum_{n}b_{n}^{\star}(t)i\hbar\frac{\partial}{\partial t}
b_{n}(t)\nonumber\\
&&+\sum_{n,m}b_{n}^{\star}(t)
\langle n|i\hbar\frac{\partial}{\partial t}|m\rangle
b_{m}(t)-\sum_{n}b_{n}^{\star}(t){\cal E}_{n}(X(t))b_{n}(t)] \}
\end{eqnarray}
where the second term in the action stands for the term
commonly referred to as Berry's phase\cite{berry} and its 
off-diagonal {\em generalization}. 
The second term in (2.12) is defined by
\begin{eqnarray} 
(U(X(t))^{\dagger}i\hbar\frac{\partial}{\partial t}U(X(t)))_{nm}
&=&\int d^{3}x v^{\star}_{n}(\vec{x},X(t))
i\hbar\frac{\partial}{\partial t}v_{m}(\vec{x},X(t))\nonumber\\
&\equiv& \langle n|i\hbar\frac{\partial}{\partial t}|m\rangle.
\end{eqnarray}
The path integral (2.12) is also derived directly by expanding $\psi(t,\vec{x})
=\sum_{n}b_{n}(t)v_{n}(\vec{x},X(t))$ in terms of  the 
instantaneous eigenfunctions in (2.10). As for the phase choice 
of $v_{n}(\vec{x},X(t))$ in (2.10), it will be discussed in 
detail later in connection with the hidden local gauge symmetry. 
As we already mentioned, the 
fact that the Berry's phase can be understood as a part of the 
Hamiltonian, i.e.,{\em dynamical}, has been noted in an 
adiabatic picture~\cite{berry2}.
Our formula does not assume the adiabatic approximation, and 
thus it gives a generalization.

In the operator formulation of the second quantized theory,
we thus obtain the effective Hamiltonian (depending on Bose or 
Fermi statistics)
\begin{eqnarray}
\hat{H}_{eff}(t)&=&\sum_{n}\hat{b}_{n}^{\dagger}(t)
{\cal E}_{n}(X(t))\hat{b}_{n}(t)\nonumber\\
&&-\sum_{n,m}\hat{b}_{n}^{\dagger}(t)
\langle n|i\hbar\frac{\partial}{\partial t}|m\rangle
\hat{b}_{m}(t)
\end{eqnarray}
with 
\begin{eqnarray}
[\hat{b}_{n}(t), \hat{b}^{\dagger}_{m}(t)]_{\mp}=\delta_{n,m}.
\end{eqnarray}
Note that these formulas (2.6), (2.12) and (2.14) are exact and, 
to our knowledge, the formulas (2.12) and (2.14) have not been 
analyzed before~\footnote{It is possible to write the 
Schr\"{o}dinger equation in the first quantization in a form
equivalent to (2.14) by expanding the Schr\"{o}dinger 
amplitude $\psi(t,\vec{x})=\sum_{n}b_{n}(t)v_{n}(\vec{x},X(t))$ 
in terms of the instantaneous 
eigenfunctions in (2.10); one then deals with simultaneous 
equations for the variables $\{b_{n}(t) \}$. However, the 
second quantization 
provides a natural universal formulation for both of the path 
integral and the operator formalism.}. See, however, eq.(2)
in ref.\cite{anandan}. The off-diagonal 
geometric terms in (2.14), 
which are crucial in the analysis below, are missing in the 
usual adiabatic approximation in the first quantization. The use 
of the instantaneous eigenfunctions in (2.12) is a common 
feature shared with the adiabatic approximation. In our picture,
 all the information about geometric phases  is included in 
the effective Hamiltonian, and for this reason we use the 
terminology ``geometric terms'' for those general terms 
appearing in the Hamiltonian. The ``geometric phases'' are used
when these terms are interpreted as phase factors of a specific
state vector.

Since our formulation starts with the path integral 
representation (2.2), the equivalence of the present exact 
formulation to the more conventional representation is expected.
It may however be nice to check this equivalence explicitly. 
We define the ``Schr\"{o}dinger'' picture by noting the 
Heisenberg equation of motion 
\begin{eqnarray}
&&i\hbar\frac{\partial}{\partial t}\hat{b}_{n}(t)=
[\hat{b}_{n}(t), \hat{H}_{eff}(t)]
\end{eqnarray}
and thus introducing a unitary operator $U(t)$ by
\begin{eqnarray}
&&i\hbar\frac{\partial}{\partial t}U(t)= - \hat{H}_{eff}(t)U(t)
\end{eqnarray}
with $U(0)=1$.
We then have 
\begin{eqnarray}
\hat{b}_{n}(t)&=&U(t)\hat{b}_{n}(0)U(t)^{\dagger},\nonumber\\
\hat{{\cal H}}_{eff}(t)&\equiv& 
U(t)^{\dagger}\hat{H}_{eff}(t)U(t)
\nonumber\\
&=&\sum_{n}\hat{b}_{n}^{\dagger}(0)
{\cal E}_{n}(X(t))\hat{b}_{n}(0)
-\sum_{n,m}\hat{b}_{n}^{\dagger}(0)
\langle n|i\hbar\frac{\partial}{\partial t}|m\rangle
\hat{b}_{m}(0).
\end{eqnarray}
We note that the state vectors in the Heisenberg and 
Schr\"{o}dinger pictures are related by 
\begin{eqnarray}
\Psi_{H}(0)=U(t)\Psi_{S}(t)
\end{eqnarray}
and thus 
\begin{eqnarray}
i\hbar\frac{\partial}{\partial t}\Psi_{S}(t)
=U^{\dagger}(t)\hat{H}_{eff}(t)U(t)U^{\dagger}(t)\Psi_{H}(0)
=\hat{{\cal H}}_{eff}(t)\Psi_{S}(t).
\end{eqnarray}

The second quantization formula for the evolution operator then 
gives rise to

\begin{eqnarray}
&&\langle n|T^{\star}\exp\{-\frac{i}{\hbar}\int_{0}^{T}
\hat{{\cal H}}_{eff}(t)
dt\}|n\rangle \nonumber\\
&=&\langle n|T^{\star}\exp\{-\frac{i}{\hbar}\int_{0}^{T}dt
[\sum_{n}\hat{b}_{n}^{\dagger}(0){\cal E}_{n}(X(t))\hat{b}_{n}(0)
-\sum_{n,m}\hat{b}_{n}^{\dagger}(0)
\langle n|i\hbar\frac{\partial}{\partial t}|m\rangle
\hat{b}_{m}(0)] \}|n\rangle
\nonumber\\
&=&\sum_{n_{1},n_{2}, ....,n_{N}}\nonumber\\
&&\langle n|\exp\{-\frac{i\epsilon}{\hbar}
[\sum_{n}\hat{b}_{n}^{\dagger}(0){\cal E}_{n}(X(T))\hat{b}_{n}(0)
-\sum_{n,m}\hat{b}_{n}^{\dagger}(0)
\langle n|i\hbar\frac{\partial}{\partial T}|m\rangle
\hat{b}_{m}(0)] \}|n_{1}\rangle
\nonumber\\
&\times&\langle n_{1}|\exp\{-\frac{i\epsilon}{\hbar}
[\sum_{n}\hat{b}_{n}^{\dagger}(0){\cal E}_{n}(X(t_{1}))
\hat{b}_{n}(0)
-\sum_{n,m}\hat{b}_{n}^{\dagger}(0)
\langle n|i\hbar\frac{\partial}{\partial t_{1}}|m\rangle
\hat{b}_{m}(0)] \}|n_{2}\rangle
\nonumber\\
&\times&
\langle n_{2}|\exp\{-\frac{i\epsilon}{\hbar}
[\sum_{n}\hat{b}_{n}^{\dagger}(0){\cal E}_{n}(X(t_{2}))
\hat{b}_{n}(0)
-\sum_{n,m}\hat{b}_{n}^{\dagger}(0)
\langle n|i\hbar\frac{\partial}{\partial t_{2}}|m\rangle
\hat{b}_{m}(0)] \}|n_{3}\rangle
\nonumber\\
&\times& ..... \nonumber\\
&\times&
\langle n_{N}|\exp\{-\frac{i\epsilon}{\hbar}
[\sum_{n}\hat{b}_{n}^{\dagger}(0){\cal E}_{n}(X(t_{N}))
\hat{b}_{n}(0)
-\sum_{n,m}\hat{b}_{n}^{\dagger}(0)
\langle n|i\hbar\frac{\partial}{\partial t_{N}}|m\rangle
\hat{b}_{m}(0)] \}|n\rangle \nonumber\\
\end{eqnarray}
where $T^{\star}$ stands for the time ordering operation and 
$\epsilon=T/(N+1)$, and the state vectors in the second 
quantization are defined by 
\begin{eqnarray}
|n\rangle=\hat{b}_{n}^{\dagger}(0)|0\rangle.
\end{eqnarray}
This formula is re-written as 
\begin{eqnarray}
&&\sum_{n_{1},n_{2}, ....,n_{N}}
[\exp\{-\frac{i\epsilon}{\hbar}[
{\cal E}_{n}(X(T))-\langle n|i\hbar\frac{\partial}{\partial T}|n_{1}\rangle]\}\delta_{n,n_{1}}
+\frac{i\epsilon}{\hbar}
\langle n|i\hbar\frac{\partial}{\partial T}|n_{1}\rangle
|_{n\neq n_{1}}]
\nonumber\\
&\times&[\exp\{-\frac{i\epsilon}{\hbar}[
{\cal E}_{n_{1}}(X(t_{1}))-\langle n_{1}|i\hbar
\frac{\partial}{\partial t_{1}}|n_{2}\rangle]\}\delta_{n_{1},n_{2}}
+\frac{i\epsilon}{\hbar}
\langle n_{1}|i\hbar\frac{\partial}{\partial t_{1}}|n_{2}\rangle
|_{n_{1}\neq n_{2}}] 
\nonumber\\
&\times&
[\exp\{-\frac{i\epsilon}{\hbar}[
{\cal E}_{n_{2}}(X(t_{2}))-\langle n_{2}|i\hbar
\frac{\partial}{\partial t_{2}}|n_{3}\rangle]\}\delta_{n_{2},n_{3}}
+\frac{i\epsilon}{\hbar}
\langle n_{2}|i\hbar\frac{\partial}{\partial t_{2}}|n_{3}\rangle
|_{n_{2}\neq n_{3}}]
\nonumber\\
&\times& ..... \nonumber\\
&\times&
[\exp\{-\frac{i\epsilon}{\hbar}[
{\cal E}_{n}(X(t_{N}))-\langle n_{N}|i\hbar\frac{\partial}
{\partial 
t_{N}}|n\rangle] \}\delta_{n_{N},n}+\frac{i\epsilon}{\hbar}
\langle n_{N}|i\hbar\frac{\partial}{\partial t_{N}}|n\rangle
|_{n_{N}\neq n}] 
\end{eqnarray}
where the state vectors in this last expression stand for the 
first quantized states defined by
\begin{eqnarray}
\hat{H}(\hat{\vec{p}}, \hat{\vec{x}},  
X(t))|n(t)\rangle
={\cal E}_{n}(X(t))|n(t)\rangle,
\end{eqnarray}
and those state vectors also appear in the definition of 
geometric terms. 
If one retains only the diagonal elements in this formula 
(2.23), one
recovers the conventional adiabatic formula~\cite{kuratsuji}
\begin{eqnarray}
\exp\{-\frac{i}{\hbar}\int_{0}^{T}dt 
[{\cal E}_{n}(X(t))
-\langle n|i\hbar\frac{\partial}{\partial t}|n\rangle]\}.
\end{eqnarray}

On the other hand,
if one retains the off-diagonal elements also, one obtains the 
exact evolution operator. We first observe, for example,
\begin{eqnarray}
&&\exp\{-\frac{i\epsilon}{\hbar}[
{\cal E}_{n_{1}}(X(t_{1}))-\langle n_{1}|i\hbar
\frac{\partial}{\partial t_{1}}|n_{2}\rangle]\}
\delta_{n_{1},n_{2}}
+\frac{i\epsilon}{\hbar}
\langle n_{1}|i\hbar\frac{\partial}{\partial t_{1}}|n_{2}\rangle
|_{n_{1}\neq n_{2}}
\nonumber\\
&=&\exp\{-\frac{i\epsilon}{\hbar}{\cal E}_{n_{1}}(X(t_{1}))\}
\langle n_{1}(t_{1})|n_{2}(t_{1}-\epsilon)\rangle 
+ O(\epsilon^{2})\nonumber\\
&=&\langle n_{1}(t_{1})|\exp\{-\frac{i\epsilon}{\hbar}
\hat{H}(\hat{\vec{p}}, \hat{\vec{x}},  
X(t_{1}))\}
|n_{2}(t_{1}-\epsilon)\rangle 
+ O(\epsilon^{2}).
\end{eqnarray}
By letting $\epsilon\rightarrow 0$, we thus obtain 
\begin{eqnarray}
&&\langle n|T^{\star}\exp\{-\frac{i}{\hbar}\int_{0}^{T}
\hat{{\cal H}}_{eff}(t)
dt\}|n\rangle\nonumber\\ 
&&=
\langle n(T)|T^{\star}\exp\{-\frac{i}{\hbar}\int_{0}^{T}
\hat{H}(\hat{\vec{p}}, \hat{\vec{x}},  
X(t))dt \}|n(0)\rangle \,.
\end{eqnarray}
Both-hand sides of this formula are exact, but the difference is 
that the geometric terms, both of diagonal and off-diagonal, 
are explicit in the second quantized formulation on the 
left-hand side. 

Here we would like to comment on the possible
advantages of using the second  quantization technique. As we
have already mentioned, all the results of the second quantization are
in principle reproduced by the first quantization in the present
single-particle problem. This fact is exemplified by the relation
(2.27). The possible advantages are thus mainly technical and 
conceptual ones. First of all, the general geometric terms are 
explicitly and neatly formulated by the second quantization both 
for the path integral (2.12) and the operator formalism (2.27). 
Also, our emphasis is on the diagonalization of the Hamiltonian 
rather than on the subtle notion of phases. This emphasis on the
 Hamiltonian is also manifest in the second quantization on the 
left-hand side of (2.27). Another technical advantage in the 
present formulation  is related to the phase freedom of the 
basis set in (2.10). The path integral formula (2.12) is based 
on the expansion
\begin{eqnarray}
\psi(t,\vec{x})=\sum_{n}b_{n}(t)v_{n}(\vec{x},X(t)),
\end{eqnarray}
and the starting path integral (2.2) depends only on the field
variable $\psi(t,\vec{x})$, not on  $\{ b_{n}(t)\}$
and $\{v_{n}(\vec{x},X(t))\}$ separately. This fact shows that 
our formulation contains a hidden local gauge symmetry 
\begin{eqnarray}
v_{n}(\vec{x},X(t))\rightarrow v^{\prime}_{n}(\vec{x},X(t))=
e^{i\alpha_{n}(t)}v_{n}(\vec{x},X(t))
, \ \ \ \  b_{n}(t) \rightarrow b^{\prime}_{n}(t)=
e^{-i\alpha_{n}(t)}b_{n}(t)
\end{eqnarray}
where the gauge parameter $\alpha_{n}(t)$ is a general 
function of $t$. One can confirm that both of the path integral measure and the action in (2.12) are invariant under this gauge transformation. By using this gauge freedom, one can choose the 
phase convention of the basis set $\{v_{n}(\vec{x},X(t))\}$ such 
that the analysis of geometric phases becomes most transparent
; in (3.4) later, we choose the basis set $\{v_{n}(y(t))\}$ such that the artificial singularity introduced by the use of 
polar coordinates becomes minimum. The meaning of this gauge 
transformation shall be explained further in connection with 
equations (3.12) and (3.21).

The expression on the right-hand side of (2.27) stands for the first 
quantized formula which has an exact path integral 
representation given by
\begin{eqnarray}
&&\langle n(T)|T^{\star}\exp\{-\frac{i}{\hbar}\int_{0}^{T}
\hat{H}(\hat{\vec{p}}, \hat{\vec{x}},  
X(t))dt \}|n(0)\rangle
\nonumber\\
&=& \iint d^{3}x(T) d^{3}x(0) u_{n}^{\star}(\vec{x}(T))
u_{n}(\vec{x}(0))\nonumber\\
&&\times 
\langle \vec{x}(T)|T^{\star}\exp\{-\frac{i}{\hbar}\int_{0}^{T}
\hat{H}(\hat{\vec{p}}, \hat{\vec{x}}, 
X(t))dt \}|\vec{x}(0)\rangle
\end{eqnarray}
and 
\begin{eqnarray}
&&\langle \vec{x}(T)|T^{\star}\exp\{-\frac{i}{\hbar}\int_{0}^{T}
\hat{H}(\hat{\vec{p}}, \hat{\vec{x}},  
X(t))dt \}|\vec{x}(0)\rangle
\nonumber\\
&=&\int_{\vec{x}(0)}^{\vec{x}(T)}{\cal D}\vec{x} {\cal D}\vec{p}
\exp\{\frac{i}{\hbar}\int_{0}^{T}dt [\vec{p}\cdot\dot{\vec{x}}-
H(\vec{p},\vec{x},X(t))] \}
\nonumber\\
&=&\int_{\vec{x}(0)}^{\vec{x}(T)}{\cal D}\vec{x} {\cal D}\vec{p}
\exp\{\frac{i}{\hbar}\int_{0}^{T}dt [\vec{p}\cdot\dot{\vec{x}}-
H(\vec{p},\vec{x},0)-\sum_{l}H_{l}(\vec{p},\vec{x})X_{l}(t)] \}
\end{eqnarray}
where the last expression is valid for sufficiently small 
$X(t)=(X_{1}(t), X_{2}(t), ... )$. In the 
analysis of level crossing, it is convenient to assume that the 
specific level crossing we are interested in takes place at the 
origin of $X(t)$ with
\begin{eqnarray}
H_{l}(\vec{p},\vec{x})=\frac{\partial H(\vec{p},\vec{x},X(t))}
{\partial X_{l}(t)}|_{X(t)=0}.
\end{eqnarray}
We note that the path integral (2.31) shows no obvious
singular behavior at the level crossing point $X(t)=0$.

\section{Level crossing and geometric  phases}

We are mainly interested in the topological properties of 
geometric phases. To simplify the analysis, we now assume that 
the level crossing takes place only between 
the lowest two levels, and we consider the familiar idealized
model with only the lowest two levels. This simplification is 
expected to be valid to analyze the topological properties 
in the  infinitesimal neighborhood of level crossing. 
The effective Hamiltonian to be analyzed 
in the path integral (2.6) is then defined  by the $2\times 2$
matrix $ h(X(t))=\left(E_{nm}(X(t))\right)$.
If one assumes that the level crossing takes place at the 
origin of the parameter space $X(t)=0$, one needs to analyze
the matrix
\begin{eqnarray}
h(X(t)) = \left(E_{nm}(0)\right) + 
\left(\frac{\partial}{\partial X_{k}}E_{nm}(X)|_{X=0}\right)
X_{k}(t)
\end{eqnarray}
 for sufficiently small $(X_{1}(1),X_{2}(1), ... )$. By a time 
independent unitary transformation, which does not induce 
an extra geometric term, the first term is diagonalized.
In the present approximation, essentially the four dimensional 
sub-space of the parameter space is relevant, and after a 
suitable re-definition of the parameters by taking linear 
combinations of  $X_{k}(t)$, we write the matrix as~\cite{berry}
\begin{eqnarray}
h(X(t))
&=&\left(\begin{array}{cc}
            E(0)+y_{0}(t)&0\\
            0&E(0)+y_{0}(t)
            \end{array}\right)
        +g \sigma^{l}y_{l}(t)\nonumber\\
\end{eqnarray}
where $\sigma^{l}$ stands for the Pauli matrices, and $g$ is a 
suitable (positive) coupling constant. This parametrization in 
terms of the variables $y_{l}$ is valid beyond the linear 
approximation, but the two-level approximation is expected to 
be valid only near the level crossing point.
 
The above matrix is diagonalized in the standard way as 
\begin{eqnarray} 
h(X(t))v_{\pm}(y)=(E(0)+y_{0}(t) \pm g r)v_{\pm}(y)
\end{eqnarray}
where $r=\sqrt{y^{2}_{1}+y^{2}_{2}+y^{2}_{3}}$  and
\begin{eqnarray}
v_{+}(y)=\left(\begin{array}{c}
            \cos\frac{\theta}{2}e^{-i\varphi}\\
            \sin\frac{\theta}{2}
            \end{array}\right), \ \ \ \ \ 
v_{-}(y)=\left(\begin{array}{c}
            \sin\frac{\theta}{2}e^{-i\varphi}\\
            -\cos\frac{\theta}{2}
            \end{array}\right)
\end{eqnarray}
by using the polar coordinates, 
$y_{1}=r\sin\theta\cos\varphi,\ y_{2}=r\sin\theta\sin\varphi,
\ y_{3}=r\cos\theta$. Note that
\begin{eqnarray}
v_{\pm}(y(0))=v_{\pm}(y(T))
\end{eqnarray}
if $y(0)=y(T)$ except for $(y_{1}, y_{2}, y_{3}) = (0,0,0)$, 
and $\theta=0\ {\rm or}\ \pi$; when one analyzes the behavior
near those singular points, due care needs to be exercised.
If one defines
\begin{eqnarray} 
v^{\dagger}_{m}(y)i\frac{\partial}{\partial t}v_{n}(y)
=A_{mn}^{k}(y)\dot{y}_{k}
\end{eqnarray}
where $m$ and $n$ run over $\pm$,
we have
\begin{eqnarray}
A_{++}^{k}(y)\dot{y}_{k}
&=&\frac{(1+\cos\theta)}{2}\dot{\varphi}
\nonumber\\
A_{+-}^{k}(y)\dot{y}_{k}
&=&\frac{\sin\theta}{2}\dot{\varphi}+\frac{i}{2}\dot{\theta}
=(A_{-+}^{k}(y)\dot{y}_{k})^{\star}
,\nonumber\\
A_{--}^{k}(y)\dot{y}_{k}
&=&\frac{1-\cos\theta}{2}\dot{\varphi}.
\end{eqnarray}
The effective Hamiltonian (2.14) is then given by 
\begin{eqnarray}
\hat{H}_{eff}(t)&=&(E(0)+y_{0}(t) + g r(t))\hat{b}^{\dagger}_{+}
\hat{b}_{+}
\nonumber\\
&+&(E(0)+y_{0}(t) - g r(t))\hat{b}^{\dagger}_{-}\hat{b}_{-}
 -\hbar \sum_{m,n}\hat{b}^{\dagger}_{m}A^{k}_{mn}(y)\dot{y}_{k}
\hat{b}_{n}.
\end{eqnarray}

In the conventional adiabatic approximation, one approximates
the effective Hamiltonian (3.8) by
\begin{eqnarray}
\hat{H}_{eff}(t)&\simeq& (E(0)+y_{0}(t) + g r(t))
\hat{b}^{\dagger}_{+}\hat{b}_{+}\nonumber\\
&&+(E(0)+y_{0}(t) - g r(t))\hat{b}^{\dagger}_{-}\hat{b}_{-}
\nonumber\\
&&-\hbar [\hat{b}^{\dagger}_{+}A^{k}_{++}(y)\dot{y}_{k}
\hat{b}_{+}
+\hat{b}^{\dagger}_{-}A^{k}_{--}(y)\dot{y}_{k}\hat{b}_{-}]
\end{eqnarray}
which is valid for 
\begin{eqnarray}
Tg r(t)\gg \hbar\pi,\nonumber
\end{eqnarray}
where $\hbar\pi$ stands for the magnitude of the geometric term 
times $T$.
The Hamiltonian for $b_{-}$, for example, is then eliminated by 
a ``gauge transformation''
\begin{eqnarray}
b_{-}(t)=
\exp\{-(i/\hbar)\int_{0}^{t}dt[
E(0)+y_{0}(t) - g r(t) 
-\hbar A^{k}_{--}(y)\dot{y}_{k}] \} \tilde{b}_{-}(t)
\end{eqnarray}
in the path integral (2.12) with the above approximation (3.9), 
and the amplitude 
$\langle 0|\hat{\psi}(T)\hat{b}^{\dagger}_{-}(0)|0\rangle$, 
which corresponds to the probability amplitude in the first 
quantization, is given by (up to an eigenfunction 
$\phi_{E}(\vec{x})$ of 
$\hat{H}(\frac{\hbar}{i}\frac{\partial}{\partial\vec{x}},
\vec{x}, 0)$ in (2.3)) 
\begin{eqnarray}
\psi_{-}(T)&\equiv&\langle 0|\hat{\psi}(T)\hat{b}^{\dagger}_{-}(0)|0\rangle\nonumber\\
&=&\exp\{-\frac{i}{\hbar}\int_{0}^{T}dt[
E(0)+y_{0}(t) - g r(t) 
-\hbar A^{k}_{--}(y)\dot{y}_{k}] \}v_{-}(y(T))
\nonumber\\
&&\times 
\langle 0|\hat{\tilde{b}}_{-}(T)\hat{\tilde{b}}{}^{\dagger}_{-}(0)|0\rangle\nonumber\\
&=&\exp\{-\frac{i}{\hbar}\int_{0}^{T}dt[
E(0)+y_{0}(t) - g r(t) 
-\hbar A^{k}_{--}(y)\dot{y}_{k}] \}v_{-}(y(T))
\end{eqnarray}
with $\langle 0|\hat{\tilde{b}}_{-}(T)
\hat{\tilde{b}}{}^{\dagger}_{-}(0)
|0\rangle=\langle 0|\hat{\tilde{b}}_{-}(0)
\hat{\tilde{b}}{}^{\dagger}_{-}(0)
|0\rangle=1$.
For a $2\pi$ rotation in $\varphi$ with fixed $\theta$, for 
example, the geometric term  gives rise to the well-known 
factor~\footnote{If one performs the gauge transformation (2.29)
 for the bases (3.4) in the formula (3.11), one can confirm
\begin{eqnarray}
\psi_{-}(T)\rightarrow e^{i\alpha_{-}(0)}\psi_{-}(T) \nonumber
\end{eqnarray}
independently of the value of $T$, and thus the amplitude 
$\psi_{-}(T)$ relative to $\psi_{-}(0)$, 
which is the quantity of physical significance, 
is independent of the gauge transformation.  }
\begin{eqnarray}
\psi_{-}(T)=\exp\{i\pi(1-\cos\theta) \}
\exp\{-\frac{i}{\hbar}\int_{C(0\rightarrow T)}dt
[E(0)+y_{0}(t) - g r(t)] \}v_{-}(y(T))
\end{eqnarray}
by using (3.7)~\cite{berry}, and the path $C(0\rightarrow T)$
specifies the integration along the above specific closed path. 
Note that $v_{-}(y(T))=v_{-}(y(0))$ in the present choice of 
the basis set, and thus (3.12) can also be written as 
\begin{eqnarray}
\psi_{-}(T)=\exp\{i\pi(1-\cos\theta) \}
\exp\{-\frac{i}{\hbar}\int_{C(0\rightarrow T)}dt
[E(0)+y_{0}(t) - g r(t)] \}\psi_{-}(0)\nonumber
\end{eqnarray}
The correction to the formula (3.12) 
arising from the finite $1/T$ may be analyzed by an iterative
procedure~\cite{berry2}, for example. One can thus analyze the 
geometric phase in the present formulation without using the 
mathematical notions such as parallel transport and holonomy.

Another representation, which is useful to analyze the behavior
near the level crossing point, is obtained by a further unitary 
transformation
\begin{eqnarray}
\hat{b}_{m}=\sum_{n}U(\theta(t))_{mn}\hat{c}_{n}
\end{eqnarray}
where $m,n$ run over $\pm$ with
\begin{eqnarray}
U(\theta(t))=\left(\begin{array}{cc}
            \cos\frac{\theta}{2}&-\sin\frac{\theta}{2}\\
            \sin\frac{\theta}{2}&\cos\frac{\theta}{2}
            \end{array}\right),
\end{eqnarray}
and the above effective Hamiltonian (3.8) is written as
\begin{eqnarray}
\hat{H}_{eff}(t)&&= (E(0)+y_{0}(t)+gr\cos\theta)
\hat{c}^{\dagger}_{+}\hat{c}_{+}\nonumber\\
&&+(E(0)+y_{0}(t)-gr\cos\theta)\hat{c}^{\dagger}_{-}
\hat{c}_{-}\nonumber\\
&&-gr\sin\theta \hat{c}^{\dagger}_{+}\hat{c}_{-}
-gr\sin\theta \hat{c}^{\dagger}_{-}\hat{c}_{+}
-\hbar\dot{\varphi} \hat{c}^{\dagger}_{+}\hat{c}_{+}.
\end{eqnarray}
In the above unitary transformation, an extra geometric
term $-U(\theta)^{\dagger}i\hbar\partial_{t}U(\theta)$ is 
induced by the kinetic term of the path integral 
representation (2.12). One can
confirm that this extra term precisely cancels the term 
containing $\dot{\theta}$ in $\hat{b}^{\dagger}_{m}
A^{k}_{mn}(y)\dot{y}_{k}\hat{b}_{n}$ as in (3.7). 
We thus diagonalize the geometric terms in this representation.
We also note that 
$U(\theta(T))=U(\theta(0))$ if $X(T)=X(0)$ except for the 
origin, and thus the initial and final states receive the same 
transformation in scattering amplitudes. The above 
diagonalization of the geometric terms corresponds to the use
of eigenfunctions
\begin{eqnarray}
w_{m}=\sum_{n}U(\theta(t))^{\dagger}_{mn}v_{n}
\end{eqnarray}
or explicitly
\begin{eqnarray}
w_{+}=\left(\begin{array}{c}
            e^{-i\varphi}\\
            0
            \end{array}\right), \ \ \ \ \ 
w_{-}=\left(\begin{array}{c}
            0\\
            1
            \end{array}\right)
\end{eqnarray}
in the definition of geometric terms.
In the infinitesimal neighborhood of the level crossing point,
namely, for sufficiently close to the origin of the parameter 
space $(y_{1}(t), y_{2}(t), y_{3}(t) )$ but 
$(y_{1}(t), y_{2}(t), y_{3}(t))\neq (0,0,0)$, one may 
approximate (3.15) by
\begin{eqnarray}
\hat{H}_{eff}(t)&\simeq& (E(0)+y_{0}(t)+gr\cos\theta)
\hat{c}^{\dagger}_{+}\hat{c}_{+}\nonumber\\
&+&(E(0)+y_{0}(t)-gr\cos\theta)\hat{c}^{\dagger}_{-}\hat{c}_{-}
-\hbar\dot{\varphi} \hat{c}^{\dagger}_{+}\hat{c}_{+}.
\end{eqnarray}
To be precise, for any given {\em fixed} time interval $T$,
\begin{eqnarray}
T\hbar\dot{\varphi}\sim 2\pi\hbar
\end{eqnarray}
which is invariant under the uniform scale transformation 
$y_{k}(t)\rightarrow 
\epsilon y_{k}(t)$. On the other hand, 
one has $T gr\sin\theta \rightarrow T\epsilon gr\sin\theta$
by the above scaling, and thus one can choose 
\begin{eqnarray}
T\epsilon gr\ll \hbar.
\nonumber
\end{eqnarray}
 The terms $\pm gr\cos\theta$ in (3.18)
may also be ignored in the present approximation.

In this new basis (3.18), the geometric phase appears only for
 the mode $\hat{c}_{+}$ which gives rise to a phase factor
\begin{eqnarray}
\exp\{i\int_{C} \dot{\varphi}dt \}=\exp\{2i\pi \}=1,
\end{eqnarray}
and thus no physical effects. In the infinitesimal neighborhood 
of level crossing, the states spanned by 
$(\hat{b}_{+},\hat{b}_{-})$ are transformed to a linear 
combination of the 
states spanned by $(\hat{c}_{+},\hat{c}_{-})$, which give no 
non-trivial 
geometric phases. The geometric terms are topological 
in the sense that they are invariant under the uniform scaling 
of $y_{k}(t)$, but their physical implications in conjunction 
with 
other terms in the effective Hamiltonian are not. For example, 
starting with the state
$\hat{b}^{\dagger}_{-}(0)|0\rangle$ one may first make 
$r\rightarrow small$ with fixed $\theta$ and $\varphi$, 
then make a $2\pi$ rotation in $\varphi$ in the bases 
$\hat{c}^{\dagger}_{\pm}|0\rangle$, and then come back to 
the original $r$ with fixed $\theta$ and $\varphi$ for a given
fixed  $T$ as in Fig.1 ; in this cycle, one does not pick up any 
non-trivial geometric phase even though one covers the solid 
angle $2\pi(1-\cos\theta)$. 
\begin{figure}[!htb]
 \begin{center}
    \includegraphics[width=10.9cm]{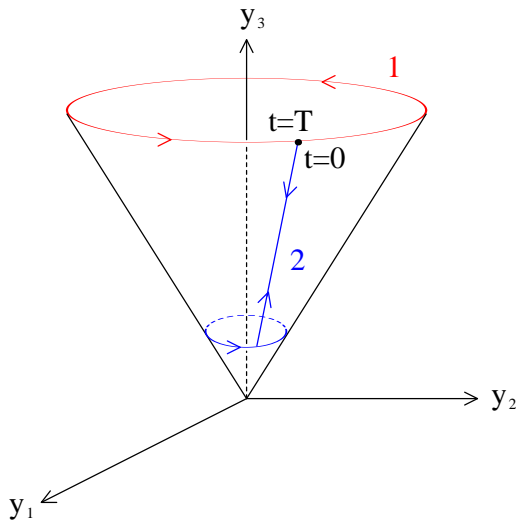} 
       \end{center}
\vspace{-9mm}      
 \caption{\small (Color online) The path 1 gives the conventional 
geometric phase as in (3.12) for a fixed finite $T$, 
whereas the path 2 gives a trivial phase  for a fixed finite $T$. 
Note that both of the paths cover the same solid angle $2\pi(1-\cos\theta)$.} 
\end{figure}

\vspace{1mm}

To be precise, the physical quantity in 
(3.12) is replaced by 
\begin{eqnarray}
\psi_{-}(T)&=&\exp\{-\frac{i}{\hbar}\int_{C_{2}(0\rightarrow T)}
dt[E(0)+y_{0}(t) - g r(t) 
-\hbar A^{k}_{--}(y)\dot{y}_{k}] \}v_{-}(y(T))\nonumber\\
&=&\exp\{-\frac{i}{\hbar}\int_{C_{2}(0\rightarrow T)}
dt[E(0)+y_{0}(t) - g r(t)] \}v_{-}(y(T))
\end{eqnarray}
by deforming the path 1 to the path 2 in the parameter space in 
Fig. 1. The path $C_{2}(0\rightarrow T)$ specifies the path 2 in 
Fig.1, and $v_{-}(y(T))=v_{-}(y(0))=\psi_{-}(0)$ in the present specific 
choice of the basis set. The first expression in the above 
equation explicitly shows the invariance of $\psi_{-}(T)$ under 
the gauge transformation (2.29) up to a trivial
overall constant phase~\footnote{The gauge transformation (2.29)
 for the present case (3.4) is written as 
\begin{eqnarray}
U(\alpha(t))=\left(\begin{array}{cc}
            e^{i\alpha_{+}(t)}&0\\
            0&e^{i\alpha_{-}(t)}
            \end{array}\right).\nonumber
\end{eqnarray}
It is convenient to keep the auxiliary variables $\{c_{m} \}$ 
and $\{w_{m} \}$ in the standard form as in (3.15) and (3.17)
even after the gauge transformation.
This is achieved by replacing $U(\theta(t))$ in (3.14) by
$U(\alpha(t))U(\theta(t))$. The effect of the gauge 
transformation survives only in the external states 
$\hat{b}_{\pm}(0)|0\rangle$ in (3.11) resulting in the 
appearance of  trivial overall constant phase.    
}. The transformation from $\hat{b}_{\pm}$ to 
$\hat{c}_{\pm}$ is highly non-perturbative, since a complete 
re-arrangement of two levels is involved.

It should be noted that one cannot simultaneously diagonalize 
the conventional energy eigenvalues and the induced geometric 
terms in (3.8) which is exact in the present two-level model 
(3.2). The topological considerations~\cite{stone, berry} are 
thus inevitably approximate. In this respect, it may be 
instructive to consider a model without level crossing which is 
defined by setting 
\begin{eqnarray}
y_{3}=\Delta E/2g
\end{eqnarray}
in (3.8), where $\Delta E$ stands for the minimum of the level 
spacing. The geometric terms then loose invariance under the 
uniform scaling of $y_{1}$ and $y_{2}$.
In the limit 
\begin{eqnarray}
\sqrt{y^{2}_{1}+y^{2}_{2}}\gg\Delta E/2g,
\end{eqnarray}
$\theta\rightarrow \pi/2$ and the geometric terms in (3.8) 
exhibit approximately topological behavior for the reduced 
variables $(y_{1},y_{2})$: One can thus perform an approximate 
topological analysis of the phase change rule.  
Near the point where the level 
spacing becomes minimum, which is specified by 
\begin{eqnarray}
(y_{1},y_{2})\rightarrow (0,0)
\end{eqnarray}
(and thus $\theta\rightarrow0$), the 
geometric terms in (3.8) assume the form of the geometric 
term in (3.18) and thus the geometric phases become trivial. 
Our analysis shows that the model 
{\em with} level crossing 
(3.2) exhibits precisely the same topological properties for any 
finite $T$. 

It is instructive to analyze an explicit 
example in Refs.~\cite{geller,bhandari} where the following 
parametrization has been introduced
\begin{eqnarray}
(y_{1},y_{2},y_{3})=(B_{0}(b_{1}+\cos\omega t), 
B_{0}\sin\omega t, B_{z})
\end{eqnarray}
and $g=\mu$ in the notation of (3.2).  The case $b_{1}=0$ and 
$B_{z}\neq 0$ corresponds to
the model without level crossing discussed above in (3.22), and 
the geometric phase becomes trivial for $B_{0}\rightarrow 0$. 
\begin{figure}[!htb]
 \begin{center}
    \includegraphics[width=10.6cm]{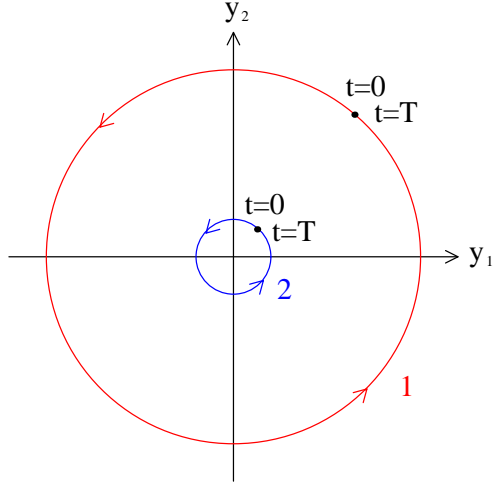} 
       \end{center}
       
\vspace{-9mm}      
 \caption{\small (Color online) The path 1 for $(y_{1},y_{2},y_{3})
=(B_{0}\cos\omega t, B_{0}\sin\omega t, 0)$ 
gives rise to the phase change rule for a fixed finite 
$T=2\pi/\omega$ and  $\mu B_{0}/\hbar\omega\gg 1$, whereas the 
path 2 gives a trivial phase for a fixed finite $T$ and 
$\mu B_{0}/\hbar\omega\ll 1$, thus resulting in the failure of 
the topological argument for the phase change rule for any fixed
 finite $T$.} 
\end{figure}

The case $b_{1}=B_{z} = 0$ describes the model with level 
crossing: The case $b_{1}=B_{z} = 0$ with
\begin{eqnarray}
T=2\pi/\omega
\end{eqnarray}
kept fixed describes the situation in (3.18) with 
$\theta=\pi/2$, namely, a closed 
cycle in the infinitesimal neighborhood of level crossing for 
$B_{0}\rightarrow 0$, and the geometric phase becomes trivial. 
See Fig.2. To be explicit, 
\begin{eqnarray}
\psi_{-}(T)&=&\exp\{-\frac{i}{\hbar}\int_{C_{1}}
dt[E(0)+y_{0}(t) - \mu B_{0}
-\hbar A^{k}_{--}(y)\dot{y}_{k}] \}v_{-}(y(T))\nonumber\\
&=&(-1)\exp\{-\frac{i}{\hbar}\int_{C_{1}}
dt[E(0)+y_{0}(t) - \mu B_{0}] \}\frac{1}{\sqrt{2}}
\left(\begin{array}{c}
            e^{-i\varphi(T)}\\
            -1
            \end{array}\right)
\end{eqnarray}
for the path 1 with $\mu B_{0}/\hbar\omega\gg 1$ where the 
factor $(-1)$ stands for the Longuet-Higgins' phase 
change~\cite{higgins}, and 
\begin{eqnarray}
\psi_{-}(T)
&=&\exp\{-\frac{i}{\hbar}\int_{C_{2}}
dt[E(0)+y_{0}(t)] \}\frac{1}{\sqrt{2}}
\left(\begin{array}{c}
            e^{-i\varphi(T)}\\
            -1
            \end{array}\right)
\end{eqnarray}
for the path 2 with $\mu B_{0}/\hbar\omega\ll 1$. Here we 
defined $v_{-}$ as a linear combination of $w_{\pm}$ in (3.17) 
to compare the result with (3.27). Note that 
$\varphi(T)=\varphi(0)$ both in (3.27) and (3.28), and thus 
\begin{eqnarray}
\psi_{-}(0)=\frac{1}{\sqrt{2}}
            \left(\begin{array}{c}
            e^{-i\varphi(T)}\\
            -1
            \end{array}\right)=\frac{1}{\sqrt{2}}
            \left(\begin{array}{c}
            e^{-i\varphi(0)}\\
            -1
            \end{array}\right).\nonumber
\end{eqnarray}

The triviality of the geometric phase persists for 
$\omega\rightarrow 0$ and  $B_{0}\rightarrow 0$ if one 
keeps    
\begin{eqnarray}
\mu B_{0}/\hbar\omega\ll 1
\end{eqnarray}
fixed for $b_{1}=B_{z} = 0$.
On the other hand, the usual adiabatic approximation (3.9) (with 
$\theta=\pi/2$ in the present model) in the neighborhood of 
level crossing is 
described by $b_{1}=B_{z} = 0$ and $B_{0}\rightarrow 0$ with 
\begin{eqnarray}
\mu B_{0}/\hbar\omega\gg 1
\end{eqnarray}
kept fixed (and thus $\omega=2\pi/T \rightarrow 0$), namely, the 
effective magnetic field is always strong; the topological 
proof of phase-change rule~\cite{stone} is based on the 
consideration of this case. (If one starts with 
$b_{1}=B_{z} = 0$ and 
$\omega=0$, of course, no geometric terms.) These cases in the 
approach to the level crossing $B_{0}\rightarrow 0$ are 
summarized in Fig.3. One recognizes that the geometric phase
is non-trivial only for a very narrow window of the parameter 
space $(\mu B_{0}, \hbar\omega)$ for small $B_{0}$ and for an essentially
 measure zero window in the approach to the level crossing 
$B_{0}\rightarrow 0$.
In this analysis, it is 
important to distinguish the level crossing problem from the 
motion of a spin $1/2$ particle; the wave functions (3.4) are 
single valued for a $2\pi$ rotation in $\varphi$ with fixed 
$\theta$.  
\begin{figure}[!htb]
 \begin{center}
    \includegraphics[width=10.9cm]{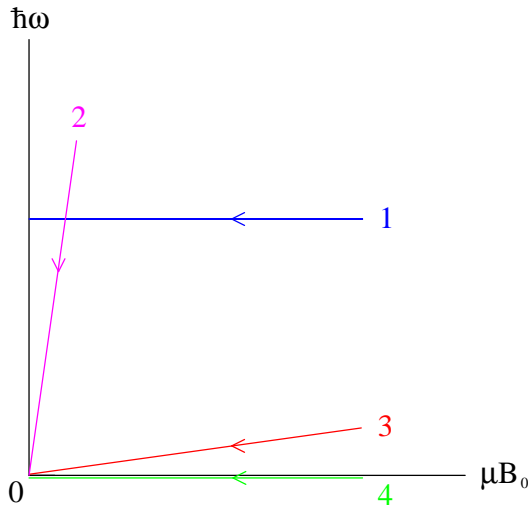} 
       \end{center}
       
\vspace{-9mm}      
 \caption{\small (Color online) Summary of the behavior of the geometric 
phases in the approach to the level crossing point $B_{0}\rightarrow 0$ in
 the parameter space $(\mu B_{0},\hbar\omega)$. The path 1 with 
fixed $\omega=2\pi/T\neq 0$ and also the path 2 with fixed 
$\mu B_{0}/\hbar\omega\ll 1$ give a trivial phase for 
$B_{0}\rightarrow 0$. The path 3 with fixed 
$\mu B_{0}/\hbar\omega\gg 1$ gives a non-trivial phase for 
$B_{0}\rightarrow 0$. The path 4 with $\omega=0$ gives no
geometric phase. The non-trivial phase arises for an essentially
measure zero set in the parameter space 
$(\mu B_{0},\hbar\omega)$ for the approach to the level crossing
 $B_{0}\rightarrow 0$. } 
\end{figure}

The conventional treatment
of geometric phases in adiabatic approximation is based on
 the premise that one 
can choose $T$ sufficiently large for any {\em given} 
$\epsilon\sim r$ such that 
\begin{eqnarray}
Tg\epsilon \gg \hbar,
\end{eqnarray}
and thus $T\rightarrow \infty$ for $\epsilon\rightarrow 0$, 
namely, it takes an infinite amount of time to approach the 
level crossing point~\cite{berry, simon}.
Finite $T$ may however be appropriate in
practical applications, as is noted in~\cite{berry}. 
Because of the uncertainty principle
$T\Delta E \geq \frac{1}{2}\hbar$,
the (physically measured) energy uncertainty for any given fixed 
$T$ is not much different from the magnitude of the geometric 
term $2\pi\hbar$, and the level spacing becomes much smaller 
than these values in the infinitesimal neighborhood of level 
crossing for the given $T$. An intuitive picture behind (3.18) is
that the motion in  $\dot{\varphi}$ smears the ``monopole'' 
singularity for arbitrarily large but finite $T$. 

In the topological analysis of the geometric phase for any fixed
finite $T$, one needs to cover the parameter regions starting 
with the region where the adiabatic approximation 
is reasonably good to the parameter region near the level 
crossing point where the adiabatic approximation totally fails.

\vspace{-2mm}

\section{Integrability of Schr\"{o}dinger equation and geometric
phase}

We here briefly comment on the integrability of Schr\"{o}dinger 
equation and the appearance of seemingly non-integrable phase
factors. The Hamiltonian (2.1), which is parametrized by a set 
of external parameters, gives rise to a unique time development
for a given $X(t)$ even in the presence of non-integrable phase 
factors. If one understands that the Hamiltonians  with  
different $X(t)$ define completely different theories, one need 
not compare theories with different  $X(t)$ and thus the issue 
of the integrability of the Schr\"{o}dinger equation does not 
directly arise. However, in the practical applications of 
geometric phases, one usually uses the Born-Oppenheimer 
approximation. The external parameters $X(t)$ then become 
dynamical variables of a more fundamental Hamiltonian, and the 
appearance of non-integrable phases suggests that one cannot 
deform  some of the paths $X(t)$ smoothly to other sets of paths
 $X(t)$. The integrability of the Schr\"{o}dinger equation 
defined by the {\it regular} fundamental Hamiltonian could then 
be spoiled, 
since the different paths $X(t)$ are supposed to be able to be 
deformed smoothly to each other for the regular Hamiltonian in 
the Schr\"{o}dinger equation. Our analysis however shows that 
geometric phases are topologically trivial for any finite time 
interval $T$ and thus the integrability of the basic
Schr\"{o}dinger equation is always ensured.

From the view point of path integral, the formula (2.6) where 
the Hamiltonian is diagonalized both at $t=0$ and $t=T$ if 
$X(T)=X(0)$ shows no obvious singular behavior at the level 
crossing point. On the other hand, the path integral (2.12) 
becomes somewhat subtle at the level crossing point; the bases 
$\{v_{n}(\vec{x},X(t))\}$ are singular on top of level crossing 
as in (3.4), and thus the unitary transformation $U$ to (2.9) 
and the induced geometric terms become singular there. The 
present analysis however shows that 
 the path integral is not singular for any finite $T$. 
This suggests that one can promote the variables $X(t)$ to 
fully dynamical variables by adding the kinetic and potential 
terms for $X(t)$,
and the path integral is still well-defined.  
We consider that this result is satisfactory since the starting 
Hamiltonian (2.1) does not contain any obvious singularity even 
when one promotes the variables $X(t)$ to fully dynamical 
variables.

\vspace{-2mm}

\section{Aharonov-Bohm phase}

It is important to clarify the similarity and difference between
the geometric phases associated with level crossing and the 
Aharonov-Bohm phase~\cite{berry, aharonov}. 
We thus start with the hermitian Hamiltonian 
\begin{equation}
\hat{H}=\hat{H}(\frac{\hbar}{i}\frac{\partial}{\partial\vec{x}},
\vec{x},A_{k}(\vec{x}))
=\frac{1}{2m}(\frac{\hbar}{i}\frac{\partial}{\partial\vec{x}}
- e\vec{A}(\vec{x}))^{2}
\end{equation}
for a single particle theory in the {\em time independent} 
background gauge potential $A_{k}(\vec{x})$
\begin{eqnarray}
&&A_{k}(\vec{x})=(-\frac{B}{2}y, \frac{B}{2}x, 0)\ \  {\rm for}
\ \ r=\sqrt{x^{2}+y^{2}} \leq a, \nonumber\\
&&A_{k}(\vec{x})=(-\frac{a^{2}B}{2r^{2}}y, \frac{a^{2}B}{2r^{2}}
x, 0)\ \  {\rm for}
\ \ r=\sqrt{x^{2}+y^{2}} \geq a
\end{eqnarray}
and thus no electric field. The uniform constant magnetic field 
$\vec{B}=\vec{\nabla}\times \vec{A}$ is confined in a cylinder 
along the $z$-axis with a radius 
$a$. The first quantized formulation of the Aharonov-Bohm effect
 is given by
\begin{eqnarray}
&&\langle \vec{x}(T)|\exp\{-\frac{i}{\hbar}\int_{0}^{T}
\hat{H}dt\}|\vec{x}(0)\rangle
\nonumber\\
&=&\int {\cal D}\vec{p}{\cal D}\vec{x} \exp\{\frac{i}{\hbar}
\int_{0}^{T} dt[\vec{p}\cdot \dot{\vec{x}}
-\frac{1}{2m}(\vec{p}-e\vec{A}(\vec{x}))^{2}] \}
\nonumber\\
&=&\int {\cal D}\vec{p}{\cal D}\vec{x} \exp\{\frac{i}{\hbar}
\int_{0}^{T} dt[(\vec{p}+e\vec{A}(\vec{x}))\cdot\dot{\vec{x}}
-\frac{1}{2m}\vec{p}\,{}^{2}] \}
\nonumber\\
&=&\int {\cal D}\vec{p}{\cal D}\vec{x} \exp\{\frac{i}{\hbar}
\int_{0}^{T} dt
[\vec{p}\cdot\dot{\vec{x}}-\frac{1}{2m}\vec{p}\,{}^{2}]
+\frac{ie}{\hbar}
\int_{C}\vec{A}(\vec{x})\cdot d\vec{x} \}
\nonumber\\
&=&\sum_n \int [{\cal D}\vec{p}{\cal D}\vec{x}]_{(n)} 
\exp\{\frac{i}{\hbar}
\int_{0}^{T} dt
[\vec{p}\cdot\dot{\vec{x}}-\frac{1}{2m}\vec{p}\,{}^{2}]
+\frac{ie}{\hbar}
n\Phi \}
\nonumber\\
&=&\sum_n \langle \vec{x}(T)|\exp\{-\frac{i}{\hbar}\int_{0}^{T}
\hat{H}_{0}dt\}|\vec{x}(0)\rangle_{(n)}
\exp\{\frac{ie}{\hbar}n\Phi \}
\end{eqnarray}
for any closed spatial path $C$, $\vec{x}(T)=\vec{x}(0)$, which 
winds the cylinder by $n$ times, and $\Phi=\int \vec{B}\cdot 
d\vec{S}$
 stands for the magnetic flux inside the cylinder. We used
the translational invariance of the path integral measure
\begin{eqnarray}
{\cal D}\vec{p}={\cal D}(\vec{p}-e\vec{A}(\vec{x}))
\end{eqnarray}
for the transformation from the second line to the third line in 
(5.3). Note that the formula (5.3) is exact, and the phase 
factor gives a truly topological quantity even for any fixed 
finite $T$; for the general case with only $\vec{x}(0)
=\vec{x}(T)$ specified, one needs to sum over $n$ in (5.3). In 
practice, the Aharonov-Bohm phase is analyzed in 
connection with interference effects, but the basic mathematical 
treatment is the same as in (5.3).

The path integral for the Aharonov-Bohm effect for the time 
interval $0\leq t\leq T$ in the second quantized 
formulation is given by 
\begin{eqnarray}
Z&=&\int{\cal D}\psi^{\star}{\cal D}\psi\nonumber\\
&\times&\exp\{\frac{i}{\hbar}\int_{0}^{T}dtd^{3}x[
\psi^{\star}(t,\vec{x})i\hbar\frac{\partial}{\partial t}
\psi(t,\vec{x})-\psi^{\star}(t,\vec{x})
\hat{H}(\frac{\hbar}{i}\frac{\partial}{\partial\vec{x}},
\vec{x},A_{k}(\vec{x}))\psi(t,\vec{x})] \}
\nonumber\\
\end{eqnarray}
We then define a complete set of eigenfunctions ( in a domain 
of 3-dimensional space with a cylinder along the $z$-axis of 
radius $a$ removed)
\begin{eqnarray}
&&\hat{H}(\frac{\hbar}{i}\frac{\partial}{\partial\vec{x}},
\vec{x},A_{k}(\vec{x}))u_{n}(\vec{x})
=E_{n}u_{n}(\vec{x}),\nonumber\\
&&\int d^{3}xu_{n}^{\star}(\vec{x})u_{m}(\vec{x})=
\delta_{nm}
\end{eqnarray}
with a suitable boundary condition on the surface of the 
cylinder and expand 
\begin{equation}
\psi(t,\vec{x})=\sum_{n}a_{n}(t)u_{n}(\vec{x}).
\end{equation}
Then
\begin{equation}
{\cal D}\psi^{\star}{\cal D}\psi=\prod_{n}{\cal D}a_{n}^{\star}
{\cal D}a_{n}
\end{equation}
and the path integral is written as 
\begin{eqnarray}
Z&=&\int \prod_{n}{\cal D}a_{n}^{\star}
{\cal D}a_{n}\nonumber\\
&\times&\exp\{\frac{i}{\hbar}\int_{0}^{T}dt[
\sum_{n}a_{n}^{\star}(t)i\hbar\frac{\partial}{\partial t}
a_{n}(t)-\sum_{n}E_{n}a_{n}^{\star}(t)a_{n}(t)] \}.
\end{eqnarray}

We next define
\begin{eqnarray}
u_{n}(\vec{x})=e^{(ie/\hbar)\int^{x}_{x(0)}A_{k}(\vec{y})dy^{k}}
v_{\vec{p}}(\vec{x})
\end{eqnarray}
and then  
\begin{eqnarray}
\frac{1}{2m}(\frac{\hbar}{i}\frac{\partial}{\partial\vec{x}}
- e\vec{A}(\vec{x}))^{2}e^{(ie/\hbar)\int^{x}_{x(0)}A_{k}dy^{k}}
v_{\vec{p}}(\vec{x})
&=&e^{(ie/\hbar)\int^{x}_{x(0)}A_{k}dy^{k}}
\frac{\hat{\vec{p}}\,{}^{2}}{2m}
v_{\vec{p}}(\vec{x})\nonumber\\
&=&E_{n}(p)u_{n}(\vec{x}),
\end{eqnarray}
namely
\begin{eqnarray}
\frac{\hat{\vec{p}}\,{}^{2}}{2m}v_{\vec{p}}(\vec{x})&=&E_{n}(p)
v_{\vec{p}}(\vec{x}),
\nonumber\\
\int d^{3}x v_{\vec{p}}^{\star}(\vec{x})
v_{\vec{p}^{\prime}}(\vec{x})&=&\delta_{\vec{p},\vec{p}^{\prime}}
\end{eqnarray}
where $v_{\vec{p}}(\vec{x})$ is defined in terms of cylindrical
coordinates (with the inside of the cylinder removed), and its 
phase convention is defined to be single valued in the sense that
\begin{eqnarray}
v_{\vec{p}}(\vec{x}(0))=v_{\vec{p}}(\vec{x}(T))
\end{eqnarray}
if $\vec{x}(0)=\vec{x}(T)$. In practical applications, one may
choose $v_{\vec{p}}(\vec{x})$ such that it approaches a plane 
wave specified by the momentum $\vec{p}$ at far away from the 
cylinder. 

Since the Hamiltonian in (5.9) is eliminated by a
 ``gauge transformation'' 
\begin{eqnarray}
a_{n}(t)=\exp\{-\frac{i}{\hbar}E_{n}t \}\tilde{a}_{n}(t),
\end{eqnarray}
we have the probability amplitude in the first quantization
\begin{eqnarray}
\langle 0|\hat{\psi}(T)\hat{a}_{n}^{\dagger}(0)|0\rangle
&=&\exp\{-\frac{i}{\hbar}E_{n}(p)T 
+ (ie/\hbar)\int^{x(T)}_{x(0)}A_{k}(\vec{y})dy^{k}\}
v_{\vec{p}}(\vec{x}(T))\nonumber\\
&&\times \langle 0|\hat{\tilde{a}}_{n}(T)
\hat{\tilde{a}}{}^{\dagger}_{n}(0)
|0\rangle
\nonumber\\
&=&\exp\{-\frac{i}{\hbar}E_{n}(p)T 
+ (ie/\hbar)\int^{x(T)}_{x(0)}A_{k}(\vec{y})dy^{k}\}
v_{\vec{p}}(\vec{x}(T))\nonumber\\
\end{eqnarray}
For a closed path $x^{k}(T)=x^{k}(0)$, we pick up the 
familiar phase factor as in (5.3).
  
Formulated in the manner (5.15), the Aharonov-Bohm phase is 
analogous to the geometric phase (3.11) associated with level
crossing, but there are several 
critical differences. First of all, the Aharonov-Bohm effect
is defined for a space which is not simply connected, and the 
Aharonov-Bohm phase is exact for any finite time interval $T$ 
(one may consider a narrow cylinder $a\rightarrow {\rm small}$ 
with the magnetic flux $\Phi=a^{2}B$ kept fixed), whereas
 the geometric phase is topologically trivial for any finite 
time interval $T$ as we have shown. The summation over the winding number $n$ in (5.3) is generally required in the case of the Aharonov-Bohm phase, but no such summation in the case of the geometric phase since the notion of the winding number is not well-defined for any fixed finite $T$.  Secondly, a closed 
path in the parameter space, which may have no direct connection 
with the real spatial coordinates, is important in the geometric
phase, whereas a closed path in the real 3-dimensional space
is important for the Aharonov-Bohm phase. Related to this last
property, the Aharonov-Bohm phase is defined for the time 
{\em independent} gauge potential, whereas the geometric phase 
is defined for the explicitly time {\em dependent} external 
parameter $X(t)$.

\section{Discussion} 

The notion of Berry's phase is known to be useful in various 
physical contexts~\cite{shapere}-\cite{review}, and the 
topological considerations are often crucial to obtain a 
qualitative understanding of what is going on. Our analysis 
however shows that the  topological interpretation of 
Berry's phase associated with level crossing generally fails in 
practical physical settings with any finite $T$. The notion of 
``approximate topology'' has no rigorous meaning, and it is 
important to keep this approximate topological property of 
geometric phases associated with level crossing in mind when one 
applies the notion of geometric phases to concrete physical 
processes. This approximate topological property  is in sharp 
contrast to the Aharonov-Bohm phase~\cite{aharonov} which is 
induced by the time-independent gauge potential and 
topologically exact for any finite time interval $T$. The 
similarity and difference between the geometric phase and the 
Aharonov-Bohm phase have been recognized in the early 
literature~\cite{berry, aharonov}, but our second quantized 
formulation, in which the analysis of the geometric phase is 
reduced to a diagonalization of the effective Hamiltonian, 
allowed us to analyze the topological properties precisely in 
the infinitesimal neighborhood of level crossing.

The correction to the geometric phase in terms of the small 
slowness parameter $\epsilon$ has been analyzed, and the closer 
to a degeneracy a system passes the slower is the necessary 
passage for adiabaticity has been noted in~\cite{berry2}. But,
to our knowledge, the fact that the geometric phase becomes
topologically trivial for practical physical settings with any 
fixed finite $T$, such as in the practical Born-Oppenheimer 
approximation where $T$ is identified with the period of the 
slower system, has been clearly stated only in the recent
paper~\cite{fujikawa}. We emphasize that this fact is 
proved independently of the adiabatic approximation. The notion 
of the geometric phase is very useful, but great care needs to 
be exercised as to its topological properties~\footnote{In page 
47 of ref.\cite{shapere}, it is stated   `` In a beautiful 1976 
paper, which the editors feel has not been  sufficiently 
appreciated,... He [A.J. Stone] showed, quite generally,  that 
the non-integrable phases imply the existence of degeneracies, 
 by means of the following topological argument."  This 
enthusiasm about  topology needs to be taken with due care.}. 

Our analysis shows that there are no mysteries about the phase 
factors of the Schr\"{o}dinger amplitude. All the information 
about the geometric phases is contained in the evolution 
operator (2.27) and thus in the path integral. The geometric 
phases are induced by the time-dependent (gauge) transformation
(2.8). One can analyze the geometric phases without referring 
to the mathematical notions such as parallel transport and 
holonomy which are useful in the framework of a precise 
adiabatic picture. Instead, the consideration of  invariance 
under the gauge symmetry (2.29) plays an important role in our
formulation. 

Also, the present path integral formulation shows a critical 
difference between the geometric phase associated with level
crossing and the quantum anomaly;
the quantum anomaly is associated with the symmetry breaking 
by the path integral measure~\cite{fujikawa2}, whereas the 
geometric phase arises from the non-anomalous terms associated 
with a change of variables as in (2.12). The similarity between 
the quantum anomaly and the geometric phase is nicely elaborated
in~\cite{jackiw}. But the quantum anomaly is basically a local 
object in the 4-dimensional space-time whereas the 
geometric phase crucially depends on the infinite time 
interval as our analysis shows. Besides, the basic symmetry
involved and its breaking mechanism in the case of geometric 
phase are not obvious. A detailed analysis of this issue will be
 given elsewhere.
\\

We thank Professor L. Stodolsky for asking if our conclusion 
is modified when the phase choice of the basis set is changed,
which prompted us to include an analysis of the hidden
local gauge symmetry into the present paper.


\begin{thebibliography}{99}
\bibitem{berry}
M.V. Berry, Proc. Roy. Soc. {\bf A392}, 45 (1984).
\bibitem{simon}
B. Simon, Phys. Rev. Lett. {\bf 51}, 2167 (1983).
\bibitem{stone}
A.J. Stone, Proc. Roy. Soc. {\bf A351}, 141 (1976).
\bibitem{higgins}
H. Longuet-Higgins, Proc. Roy. Soc. {\bf A344}, 147 (1975).
\bibitem{wilczek}
F. Wilczek and A. Zee, Phys. Rev. Lett. {\bf 52}, 2111 (1984).
\bibitem{kuratsuji}
H. Kuratsuji and S. Iida, Prog. Theor. Phys. {\bf 74}, 439 
(1985).
\bibitem{anandan}
J. Anandan and L. Stodolsky, Phys. Rev. D{\bf 35}, 2597 (1987).
\bibitem{aharonov}
Y. Aharonov and J. Anandan, Phys. Rev. Lett. {\bf 58}, 1593 
(1987).
\bibitem{berry2}
M.V. Berry, Proc. Roy. Soc. {\bf A414}, 31 (1987).
\bibitem{manini}
N. Manini and F. Pistolesi, Phys. Rev. Lett. {\bf 85}, 3067 
(2000).
\bibitem{mukunda}
N. Mukunda, Arvind, E. Ercolessi, G. Marmo, G. Morandi, and 
R. Simon, Phys. Rev. A{\bf 67}, 042114 (2003).
\bibitem{hasegawa1}
Y. Hasegawa, R. Loidl, M. Baron, G. Badurek, and H. Rauch,
Phys. Rev. Lett. {\bf 87}, 070401 (2001).
\bibitem{hasegawa2}
Y. Hasegawa, R. Loidl, G. Badurek, M. Baron, N. Manini, F. Pistolesi, and 
H. Rauch, Phys. Rev. A{\bf 65}, 052111 (2002).
\bibitem{fujikawa}
K. Fujikawa, Mod. Phys. Lett. A{\bf 20}, 335 (2005),
quant-ph/0411006.
\bibitem{geller}
Y. Lyanda-Geller, Phys. Rev. Lett. {\bf 71}, 657 (1993).
\bibitem{bhandari}
R. Bhandari, Phys. Rev. Lett. {\bf 88}, 100403 (2002). 
\bibitem{shapere}
A. Shapere and F. Wilczek, ed., {\em Geometric Phases in 
Physics} (World Scientific, Singapore, 1989), and
papers reprinted therein.
\bibitem{review}
As for a recent account of this subject see, for 
example, D. Chruscinski and A. Jamiolkowski,{\em Geometric 
Phases in Classical and Quantum Mechanics} (Birkhauser, Berlin, 
2004).
\bibitem{fujikawa2}
K. Fujikawa, Phys. Rev. Lett. {\bf 42}, 1195 (1979); Phys.
Rev. D{\bf 21}, 2848 (1980).
\bibitem{jackiw}
R. Jackiw, "Three Elaborations on Berry's Connection, Curvature
and Phase", Int. J. Mod. Phys. A {\bf 3}, 285 (1988), which is 
reprinted in the book in Ref.\cite{ shapere}.

\end{thebibliography}
\end{document}